\documentclass[aps,prl,twocolumn,superscriptaddress]{revtex4-1}

\usepackage{graphicx} 
\usepackage{amssymb} 
\usepackage{amsmath}
\usepackage{xspace}
\usepackage{longtable}
\usepackage{soul}
\usepackage{xcolor}
\usepackage{epstopdf}
\usepackage{array}
\usepackage{tabularx}
\usepackage{physics}

\usepackage[colorlinks = true,linkcolor = blue,urlcolor  = blue,citecolor = blue,anchorcolor = blue]{hyperref}

\newcommand{\HoW}{HoW$_{10}$\xspace}
\newcommand{\Efield}{$E$-field\xspace}
\newcommand{\df}{$\delta f$\xspace}

\begin{document}

\title{\textcolor{black}{Quantum coherent spin-electric control in a molecular nanomagnet at clock transitions}}
\author{Junjie Liu}
\email{junjie.liu@physics.ox.ac.uk}
\affiliation{CAESR, Department of Physics, University of Oxford, The Clarendon Laboratory, Parks Road, Oxford OX1 3PU, UK}
\author{Jakub Mrozek}
\affiliation{CAESR, Department of Physics, University of Oxford, The Clarendon Laboratory, Parks Road, Oxford OX1 3PU, UK}
\author{Aman Ullah}
\affiliation{Instituto de Ciencia Molecular (ICMol), Universitat de Val\`encia, Paterna, Spain}
\author{Yan Duan}
\affiliation{Instituto de Ciencia Molecular (ICMol), Universitat de Val\`encia, Paterna, Spain}
\author{Jos\'e J. Baldov\'i}
\affiliation{Instituto de Ciencia Molecular (ICMol), Universitat de Val\`encia, Paterna, Spain}
\author{Eugenio Coronado}
\affiliation{Instituto de Ciencia Molecular (ICMol), Universitat de Val\`encia, Paterna, Spain}
\author{Alejandro Gaita-Ari\~no}
\email{Alejandro.Gaita@uv.es}
\affiliation{Instituto de Ciencia Molecular (ICMol), Universitat de Val\`encia, Paterna, Spain}
\author{Arzhang Ardavan}
\email{arzhang.ardavan@physics.ox.ac.uk}
\affiliation{CAESR, Department of Physics, University of Oxford, The Clarendon Laboratory, Parks Road, Oxford OX1 3PU, UK}

\maketitle

{\bf Electrical control of spins at the nanoscale offers significant architectural advantages in spintronics, because electric fields can be confined over shorter length scales than magnetic fields~\cite{Kane1998,Trif2008,Laucht2015,Tosi2017,Asaad2019}. Thus, recent demonstrations of electric-field (\Efield) sensitivities in molecular spin materials~\cite{Liu2019,Fittipaldi2019,Robert2019} are tantalising, raising the viability of the quantum analogues of macroscopic magneto-electric devices~\cite{Palii2014,CardonaSerra2015,Gaita-Arino2019,Atzori2019,Godfrin2017a,Eerenstein2006,Matsukura2015}.
However, the \Efield sensitivities reported so far are rather weak, prompting the question of how to design molecules with stronger spin-electric couplings. Here we show that one path is to identify an energy scale in the spin spectrum that is associated with a structural degree of freedom with a significant electrical polarisability. We study an example of a molecular nanomagnet in which a small structural distortion establishes clock transitions (i.e.\ transitions whose energy is to first order independent of magnetic field) in the spin spectrum; the fact that this distortion is associated with an electric dipole allows us to control the clock transition energy to an unprecedented degree. We demonstrate coherent electrical control of the quantum spin state and exploit it to manipulate independently the two magnetically-identical but inversion-related molecules in the unit cell of the crystal. Our findings pave the way for the use of molecular spins in quantum technologies and spintronics.}


The polyoxometalate molecular anion [Ho(W$_5$O$_{18}$)$_2$]$^{9-}$ (abbreviated to \HoW), \textcolor{black}{within the crystal structure Na$_9$[Y$_{1-x}$Ho$_x$(W$_5$O$_{18}$)$_2$)]$\cdot$35H$_2$O ($x = 0.1\%$)}, provides an example of clock transition (CT) molecular spin qubit~\cite{AlDamen2009,Shiddiq2016}. In the solid state, the sodium salt of this anion crystallizes in a primitive space group of $P\bar{1}$, where each unit cell contains two \HoW anions that are inversion-symmetry related. Each \HoW possesses an approximate $D_{4d}$ symmetry. The magnetic properties of the \HoW, which are characterised by a total electronic angular momentum of $J=8$ and a nuclear spin of $I=7/2$, can be described by the Hamiltonian~\cite{Ghosh2012} 
\begin{equation}
\label{HoW_H}
\begin{aligned}
\hat{H} = \sum_{k = 2,4,6}\sum_{q = -k}^{k}B_k^q\hat{O}_k^q + \hat{J}\cdot A\cdot\hat{I}\\
+ \mu_\mathrm{B}g_\mathrm{e}\boldsymbol{B}_0\cdot\hat{J} - \mu_\mathrm{N}g_\mathrm{N}\boldsymbol{B}_0\cdot\hat{I}
\end{aligned}
\end{equation}
where $A$ is the magnitude of the (approximately) isotropic hyperfine interaction, $g_\mathrm{e}$ and $g_\mathrm{N}$ are the electronic and nuclear gyromagnetic ratios respectively, the anisotropy is parameterised by the amplitudes $B_k^q$ of the extended Stevens operators $\hat{O}_k^q$,
and $\boldsymbol{B}_0$ is the applied magnetic field.

The crystal field terms $\sum_{k}B_k^0\hat{O}_k^0$ lead to a ground state of $m_J = \pm 4$, where $m_J$ is the projection of the electronic angular momentum. 
Crucially, owing to interactions of \HoW with counterions and crystallisation water, there is a minor deviation from ideal $D_{4d}$ symmetry, with a continuous-shape measurement $S < 0.1$ around the Ho$^{3+}$ ion as defined by SHAPE~\cite{SI, Alvarez2005}. Specifically, the chemical structure shows that the skew angle, $\theta$, deviates from the ideal value of 45$^\circ$, and that the Ho$^{3+}$ centre deviates from the centre position by a distance of $d=(h-h')/2$. The latter indicates that the Ho$^{3+}$ is closer to one of the two [W$_5$O$_{18}$]$^{6-}$ moieties. The crystal structures measured at three different temperatures (100K, 150K, and 200K) are provided in Ref~\cite{SI} (see Section~II). At these three temperatures, the deviations of the skew angles are 2.21$^\circ$, 2.34$^\circ$ and 2.27$^\circ$, whereas the off-centre distances are 0.021~\AA, 0.025~\AA\, and 0.024~\AA, respectively.

This gives rise to an electric dipole moment and a tetragonal spin anisotropy (parameterised by terms $B_k^4$) which mixes the $m_J = \pm 4$ ground states and generates a series of anticrossings in the ground spin spectrum, leading to four electron spin resonance (ESR) CTs, whose resonance frequencies are determined by the amplitude of the tetragonal anisotropy. The four CTs correspond to resonances with different electron--nuclear spin states $\vert m_J, m_I \rangle$ with $m_J = \pm4$ and of $m_I =$ -1/2, -3/2, -5/2 or -7/2 (from low to high fields).
Previous ESR and magnetization studies~\cite{Ghosh2012,Shiddiq2016} are satisfactorily simulated by the parameters $B_2^0/h = 18.0\times10^{3}$~MHz, $B_4^0/h = 209$~MHz, $B_6^0/h = 1.53$~MHz, $B_4^4/h = 94.2$~MHz, $A/h = 830$~MHz and $g_\mathrm{e} = 1.25$. 

Thus the broken inversion symmetry causes the spin-electric coupled phenomena of an electric dipole and a clock transition. This, in turn, allows us to manipulate the CT frequency linearly by applying an external electric field (\Efield).

\begin{figure}[t]
\centering
\includegraphics[width=\columnwidth]{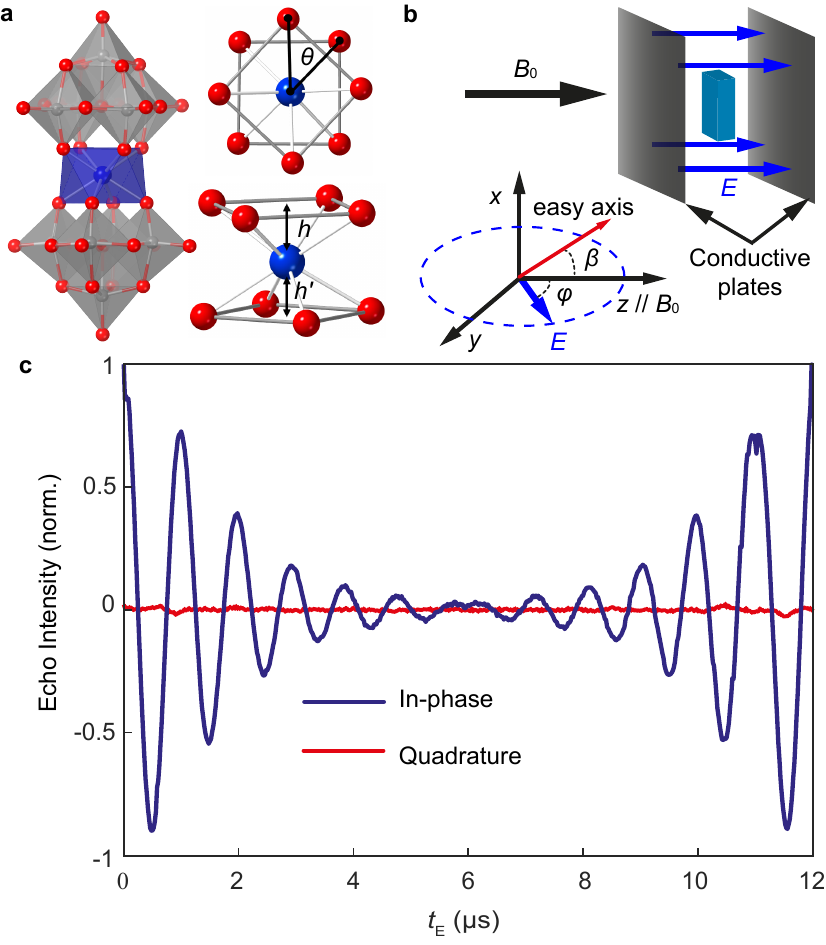}
\caption{
(a) \HoW (left) and the coordination environment for the Ho$^{3+}$ (right). A small rotation of the two Ho-coordinated [W$_5$O$_{18}$]$^{6-}$ ligands ($\theta$) and displacement of Ho ($h \neq h^\prime$) give rise to a tetragonal anisotropy. 
(b) Schematic plot showing the experiment configuration. The electric field was generated by applying voltage pulses to two parallel conducting plates. The magnetic field was applied parallel to the $z$-axis (in the $x-y-z$ laboratory  frame). 
The sample was oriented to minimize the angle between the molecular easy axis (the red arrow in the $x - z$ plane) and $\boldsymbol{B}_0$ (estimated misalignment $\beta\sim 38^\circ$). 
The \Efield orientation $\varphi$ was rotated within the $y-z$ plane. 
(c) \Efield effect on the spin echo of \HoW measured at $B_0 = 0.0304$~T. The in-phase part of the integrated echo intensity oscillates strongly as a function of the duration ($t_\mathrm{E}$) of the \Efield pulse generated by applying a voltage of 300~V. By comparison, the quadrature component (red line) remains flat at zero throughout the experiment. The durations of the $\pi/2$ and $\pi$ microwave pulses were 32 and 64~ns respectively, and the separation between them was fixed at 6~$\mu$s. All the SEC measurements presented in this work (except the selective spin excitation data shown in Fig.~\ref{fig3}) were conducted using the same protocol and the shift in the ESR frequency, $\delta f$, is obtained by Fourier transforming the oscillating in-phase component of the spin echo.
}
\label{fig1}
\end{figure}

We investigated the spin-electric coupling (SEC) in \HoW by embedding \Efield pulses in an ESR Hahn-echo sequence, as described in Refs~\cite{Liu2019,SI} and Methods.
Fig.~\ref{fig1}c shows typical data recorded with the \Efield applied parallel to $\boldsymbol{B}_0$, $\varphi=0$. The in-phase part of spin echo shows a pronounced oscillation upon varying the length of the \Efield pulse. The oscillation arises because the molecular spin Hamiltonian is modulated by the \Efield via the SEC, leading to a shift \df of the ESR frequency while the \Efield is applied. This shift in the ESR frequency manifests as an additional phase of $\delta f \cdot t_\mathrm{E}$ in the spin echo signal, causing the oscillation of period 1~$\mu$s in the echo amplitude (see Supplementary Figure S1). The decay in the echo amplitude over $0 < t_\mathrm{E} \leq 6~\mu$s is due to a small inhomogeneity in the \Efield across the crystal; during the second period of free evolution $6~\mu$s~$ < t_\mathrm{E} \leq 12~\mu$s, the inhomogeneity is refocused and the echo amplitude recovers~\cite{Liu2019}.

The lack of a dependence of the quadrature channel of the echo on $t_\mathrm{E}$ is evidence of a linear, as opposed to quadratic, SEC in \HoW. The crystal unit cell contains two \HoW units related by inversion symmetry; a linear SEC shifts the ESR frequency of each by the same amount but in opposite directions. Hence the phase shifts for the spin echoes of the two inequivalent spins are $+\delta f \cdot t_\mathrm{E}$ and $-\delta f \cdot t_\mathrm{E}$ respectively, and the quadrature part of the combined echo signal remains zero, independent of $t_\mathrm{E}$. For a second order SEC $\delta f \propto E^2$, both spins in the unit cell acquire the same phase, because the shift is insensitive to the polarity of the \Efield. Thus an oscillation in the quadrature component, $\pi/2$ out of phase with the in-phase component, is a signature of a second-order SEC~\cite{Enote}.

\begin{figure*}[t]
\centering
\includegraphics[width=1.5\columnwidth]{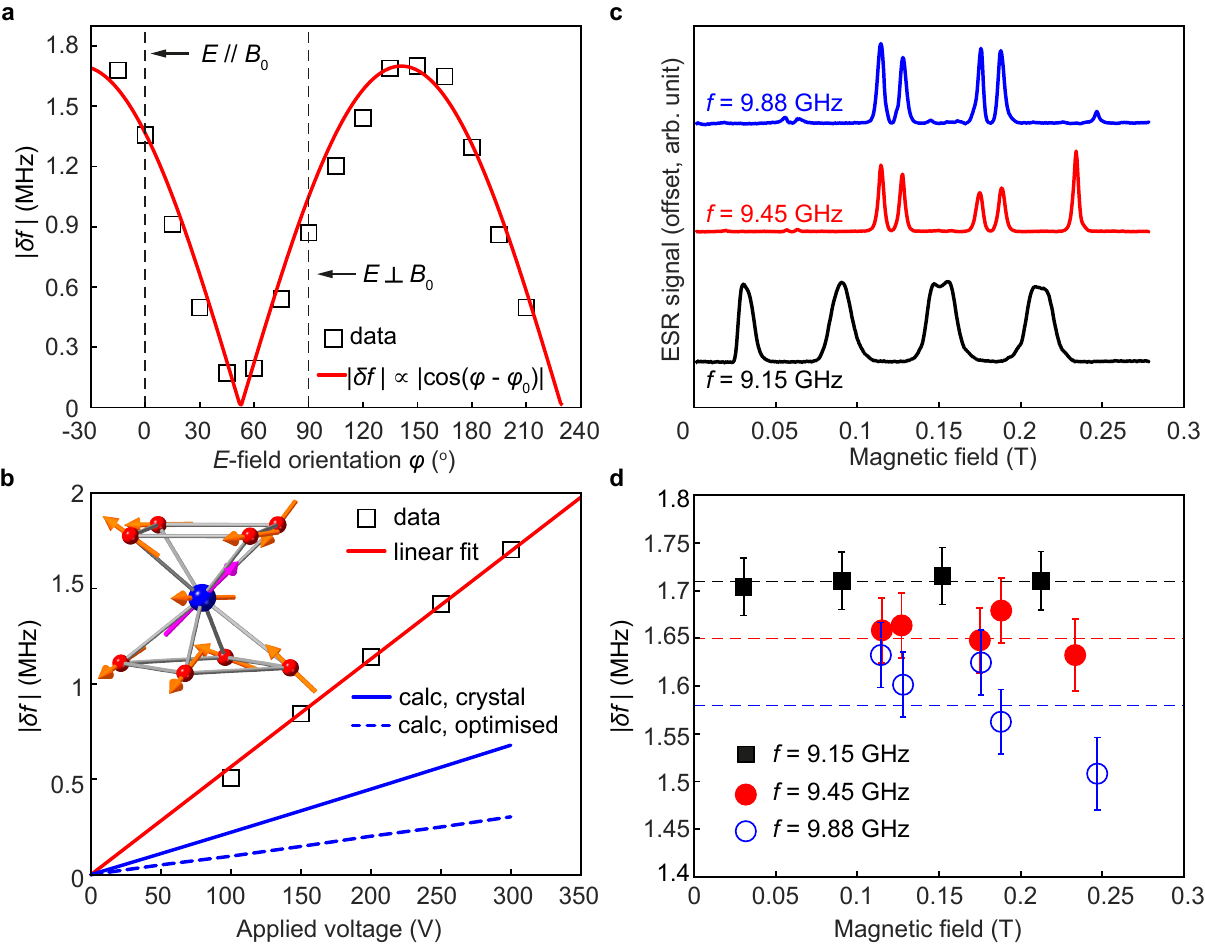}
\caption{
(a) Orientation dependence of the \Efield induced frequency shift. The measurements were performed with $f = 9.15$~GHz , $V = 300$~V and $B_0 = 0.0304$~T. The \Efield was rotated between parallel to $\boldsymbol{B}_0$ (i.e. $38^\circ$ away the molecular easy-axis) at $\varphi = 0$ (180$^\circ$) and perpendicular to $\boldsymbol{B}_0$ (i.e.\ in the molecular hard plane), $\varphi = 90^\circ$. 
(b) The measured frequency shift ($\square$) versus the applied voltage, $V$, showing a linear \Efield coupling in \HoW. The data were recorded at the orientation with the strongest SEC ($\varphi = 145^\circ$). The red line is a linear fit to the data. Blue lines represent the \textit{ab initio} prediction based on molecular geometry extracted from (solid) single-crystal X-ray crystallography or (dashed) structure fully optimsed at DFT level (see details in Ref~\cite{SI}). Inset: the local environment of the Ho showing the electric dipole direction (magenta arrow) and the \Efield-induced atomic displacement directions (orange arrows).
(c) ESR spectra recorded at various frequencies close to the CT frequency. 
(d) The \Efield effect measured on the corresponding ESR transitions. The dashed lines illustrate the \Efield-induced frequency shift \df expected if only $B_4^4$ were modified by the \Efield; the scatter in \df at frequencies away from the clock transition indicates that the \Efield-sensitivity of other Hamiltonian terms becomes important. The error bars for $\delta f$ in (a) and (b) are approximately equal to the size of the symbols (not shown).
}
\label{fig2}
\end{figure*}

The orientation dependence of the SEC is shown in Figure~\ref{fig2}a. The shift in the ESR frequency \df is calculated by taking the fast Fourier transform of the oscillation in the in-phase echo intensity. It depends on orientation as $|\delta f| \propto |\cos{\varphi - \varphi_0}|$ (the presence of both inversion-related populations means that we cannot distinguish the sign of $\delta f$ from this measurement). This yields a lower bound on the scale of the SEC; a full mapping of the SEC orientation dependence would be required to establish the orientation with maximum \Efield sensitivity. This would depend on two-axis rotation of the electric field, which is beyond the scope of this study.

The linearity of the SEC is further confirmed by varying the amplitude of the \Efield pulse of fixed duration. Figure~\ref{fig2}b demonstrates that the frequency shift is proportional to the voltage (and hence, since the electrode geometry is fixed, the amplitude of the \Efield). The data were recorded at the orientation $\varphi$ that shows the strongest SEC. The linear fit to the data yields a spin-electric coupling constant of $11.4\pm 0.3~\mathrm{Hz}/\mathrm{Vm}^{-1}$.

The relative strength of the SEC in \HoW showcases the potential for chemical design in enhancing desired molecular properties via prudent choice of the coordination environment of the metal ion. Such engineering is not possible in, for example, atomic defects in solid state materials, in which structures and therefore properties are much less tunable. Furthermore, the possibility of tuning the ESR transition at the CT fields allows the exploitation of the strong spin orbit coupling characteristic of $4f$ electrons to enhance the electrical control of molecular qubits, while retaining substantial coherence times. Such tuning of the \HoW CT can only be achieved efficiently by directly adjusting the tetragonal anisotropy interaction since, at the CT fields, the ESR transitions are insensitive to $g_\mathrm{e}$ or $A$ to first order~\cite{SI}. 

The ESR frequency at the CT fields is determined by the tetragonal anisotropy. Therefore, all ESR transitions at 9.15~GHz exhibit the same response to the applied electric fields (filled squares in Fig.~\ref{fig2}d) and can be fitted with a $\delta B_4^4/h = 8.8\pm0.2\times10^{-3}$~MHz ($5.9\times10^{-2}$~Hz/Vm$^{-1}$). However, other spin Hamiltonian terms, such as Zeeman and hyperfine interactions, may potentially also exhibit \Efield sensitivities. These interactions modify the ESR transition frequencies away from the CT fields; we can probe their sensitivities by studying the \Efield effect on ESR transitions away from the CT fields, as shown in Fig.~\ref{fig2}c and d.

The \Efield-induced ESR frequency shift decreases as the ESR frequency increases, which is expected as the ``anticrossing'' effect of $B_4^4$ reduces when moving away from the CTs. On the other hand, the SEC effects also fluctuate considerably from peak to peak at 9.45 and 9.88~GHz. Such fluctuation is likely due to the \Efield modulation of Hamiltonian parameters other than $B_4^4$: the appreciable dependence of \df on magnetic field and nuclear spin projection away from the CT fields (i.e.\ between states with varying contributions from different nuclear spin projections $m_I$ and electron spin projections $m_J$) suggest that Zeeman and hyperfine interactions may also be sensitive to the \Efield, though less so than $B_4^4$~\cite{SI}. 

We seek insight into the relationship between the \Efield-induced distortion and the CT frequency by noting that any molecular distortion may be decomposed into displacements of the normal modes of the \HoW, which we obtain using density functional theory (DFT, implemented by Gaussian~\cite{g16}). Each normal mode is associated with a force constant $\kappa_i$ and a reduced mass $\mu_i$ (yielding an eigenfrequency $\omega_i = \sqrt{\kappa_i/\mu_i}$). The electric dipole ${\boldsymbol p}$ depends on the displacement of the mode $x_i$, and this determines the coupling of the mode to an applied \Efield or to incident light, i.e.\ its infrared (IR) intensity.

When an external electric field ${\boldsymbol E}$ is incremented by ${\mathrm d}{\boldsymbol E}$, it elastically distorts the molecule by ${\mathrm d}x_i$ and modifies the molecular electric dipole by ${\mathrm d}{\boldsymbol p}_i$ for mode $i$. In this process, it does work ${\boldsymbol E}\cdot\sum_i{\mathrm d}{\boldsymbol p}_i = \sum_{i} \kappa_i\,{\mathrm d}x_i$. Thus by calculating the electric dipole moment as a function of the mode displacements, we can extract quantitatively the displacements as a function of applied \Efield,
and hence the electric-field-induced evolution of the molecular electronic structure (see Methods).

This reasoning yields criteria by which we may determine whether a particular molecular mode leads to a strong contribution to the SEC at the CT frequency: it should be relatively soft (i.e.\ exhibit small $\kappa_i$), allowing a significant molecular displacement without excessive elastic energy cost; it should couple strongly to the molecular electric dipole (i.e.\ ${\mathrm d}{\boldsymbol p}_i/{\mathrm d}x_i$, and therefore its IR intensity, should be large); and it should modify the Ho environment such that the energy of the anticrossing levels is modulated.  
Our analysis of the basis of the HoW$_{10}$ vibrational modes of the crystallographic structure reveals that the molecular displacement responsible for the SEC can be approximated by the distortion displacing the Ho and coordinating oxygen atoms as shown in the inset to Fig.~2(b). The analysis based on the relaxed structure yields a very similar result. Animations showing each of these collective distortions are available as SI files. The difference between the SECs predicted using the optimised and the crystal structures (blue dashed/solid lines in Fig.~\ref{fig2}b) can be understood by considering how the structure deviates from the ideal $D_{4d}$ symmetry. A departure from the $D_{4d}$ structure, which is inversion symmetric, is necessary to afford a linear SEC in HoW$_{10}$. Compared to the optimised molecular geometry, the actual crystal structure shows a larger distortion due to the presence of the Na$^+$ counterions and crystallisation H$_2$O molecules. Therefore, it is conceivable that the optimised structure leads to an underestimation of the SEC. Nevertheless, the theoretical results are in reasonable agreement with the experimental data.

We note in passing that, from a practical perspective, any external electric field that can be applied experimentally is very small compared to the intra-molecular fields associated with chemical bonds, justifying our perturbative approach. Others have investigated the effects of electric fields on the $\sim 10\,\mathrm{V} / \mathrm{nm}$ scale on lanthanide single-ion molecular magnets, a regime that can be explored directly in DFT~\cite{Sarkar2020}.

\begin{figure}[t]
\centering
\includegraphics[width=\columnwidth]{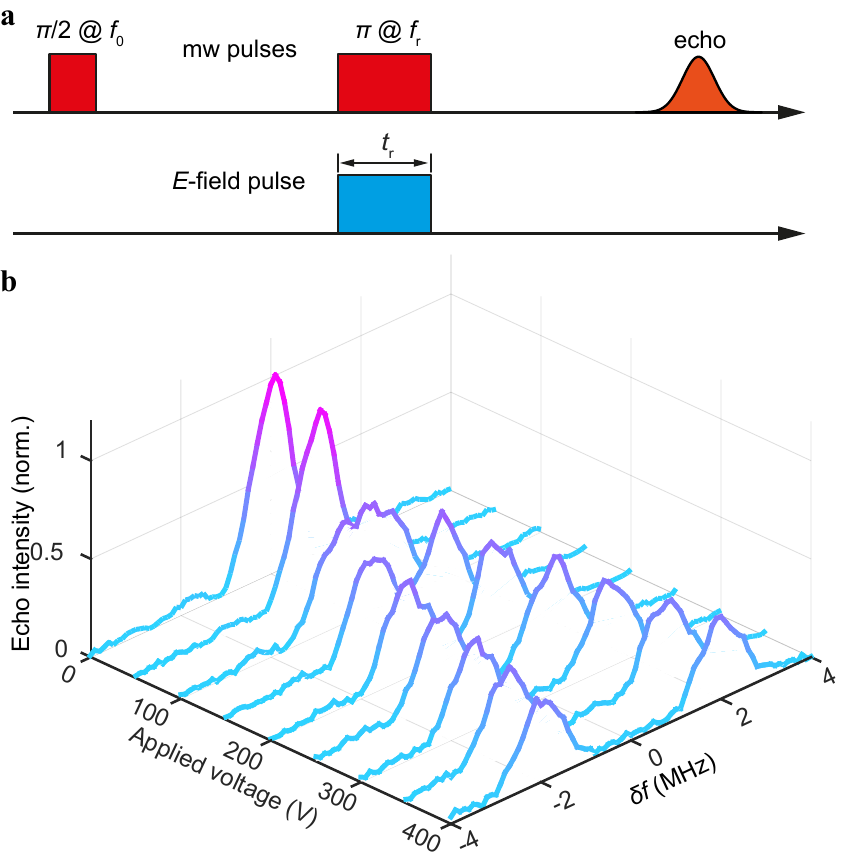}
\caption{
(a) The modified Hahn-echo pulse sequence used for selective spin excitation. A $\pi/2$ pulse is applied at the CT frequency $f_0$, while the frequency of the refocusing $\pi$ pulse, $f_{\mathrm r}$, is swept. An \Efield square pulse is applied simultaneously with the refocussing pulse to modulate the excitation frequencies of molecular spins. When $f_{\mathrm r} = f_0 \pm \delta f$, the refocusing pulse selectively inverts one or other of the inversion-symmetry-related molecular spins in the unit cell. (b) The intensity of the spin echo as a function of the applied voltage, $V$, and the frequency offset of the refocusing pulse, $\delta f=f_0-f_r$.}
\label{fig3}
\end{figure}

Finally, we demonstrate a protocol to selectively manipulate the spins of inversion-related \HoW anions assisted by an \Efield~\cite{Kane1998,Laucht2015}. Since the two spins within the unit cell are related by inversion symmetry, they are magnetically identical and cannot be  distinguished in conventional ESR or magnetometry experiments. However, they exhibit opposite frequency shifts in the presence of an \Efield, which can be exploited, using a modified Hahn-echo sequence as shown in Figure~\ref{fig3}a, to excite them selectively. The first pulse, a $\pi/2$ pulse at the CT frequency of $f_0$ = 9.15~GHz, places all molecular spins in a superposition state. An \Efield is applied simultaneously with the refocusing $\pi$ pulse, lifting the degeneracy of the ESR transitions in the two inversion-related populations. Thus it is possible to refocus only one subpopulation, as long as (i) the $\pi$ pulse is resonant with the shifted ESR frequency of that subpopulation, and (ii) the shift is larger than the natural line width of the excited spin population. The echo that forms following the second period of free evolution is detected at the frequency $f_0$, but is comprised only of the spins from the refocused subpopulation.

We set the durations of the $\pi/2$ and $\pi$ microwave pulses to be 400~ns and 800~ns respectively, to selectively excite a narrow frequency population of spins. (Note that these pulses are more than ten times longer than those used in the preceding experiments, and therefore drive a correspondingly narrower range of ESR transition frequencies.) Figure~\ref{fig3}b shows a density plot of the echo amplitude as a function of the applied voltage $V$ and the frequency offset of the refocusing pulse $\delta f=f_0-f_r$ (see section VI in Ref~\cite{SI} for more data). In the absence of an \Efield pulse ($V=0$), the maximum echo signal is observed when the frequencies of the $\pi/2$ and $\pi$ pulses are identical, i.e.\ $\delta f = 0$. Upon increasing the amplitude of the \Efield, subpopulations are refocused for two different values of \df, and these subpopulations are spectrally well-resolved for $V>150$~V. 

The offset is in excellent agreement with the expected SEC for \HoW and the echo signal peaks at the symmetric positions $\pm \delta f$ around $\delta f = 0$, corresponding to the selective refocusing of the inversion-related subpopulations. Furthermore, the echo intensities at $\pm \delta f$ are identical and approximately half of the intensity in the absence of an electric field, indicating that only half of the population is refocused, and that the refocusing is equally effective for both subpopulations. This demonstrates that, with the assistance of an electric field, we can distinguish orientations of otherwise magnetically identical \HoW units.

As is generally the case in quantum information experiments, strong coupling to a control field means strong coupling to a source of noise that shares the same physics as the control. In the case described here, we reduce sensitivity to magnetic fields by working at a CT, but this in turn increases sensitivity to molecular distortions and E-field fluctuations. We note, however, that reducing temperature (which will in any case be required for initialisation) suppresses environmental phonons (the principal source of molecular distortions), but cannot remove all magnetic field fluctuations, which may have their origin in, for example, environmental spin flip-flop processes~\cite{Wedge2012}.

Our results guide us to distill the general recipe for a high-SEC-molecule: a soft and electrically polarisable environment for the spin carriers, and a spin spectrum that is highly sensitive to distortions.
These principles are satisfied in \HoW, where a clock transition frequency is modulated by soft modes that shift the molecular charge distribution.
The SEC is about an order of magnitude larger than the values the previously reported for transition-metal-based molecular nanomagnets~\cite{Liu2019,Fittipaldi2019,Kinzel2021} ($\delta f/E < 1$~Hz/Vm$^{-1}$) and it also surpasses the SEC measured for rare earth atoms doped in YAG~\cite{liu2020} ($\delta f/E \approx 1$~Hz/Vm$^{-1}$). It is similar to the SEC for Mn$^{2+}$ in ZnO~\cite{George2013} ($\delta f/E = 12.7$ and 6.4~Hz/Vm$^{-1}$ for $m_s: \pm5/2 \leftrightarrow \pm3/2$ and $\pm3/2 \leftrightarrow \pm1/2$ transitions, respectively), in which  the SEC is associated with the piezoelectric nature of the host lattice. 
The \HoW coupling is sufficiently strong that a modest \Efield of $\sim10^{5}$~V/m or 100~$\mu$V/nm is adequate to tune the spin at a practically useful level, i.e. to shift the frequency by much more than the natural line width; this demonstrates the principle that local \Efield tuning allows for selective addressing of spins in otherwise identical molecules. The strong SEC in \HoW raises the tantalizing possibility of engineering a coherent spin-photon interface in molecular spintronic devices~\cite{Mi2017}, allowing coherent spin control by an oscillating electric field~\cite{Asaad2019}.
These results pave the way for the use of molecular components in quantum or classical spintronic technologies in which local electrical control can surpass the performance of conventional magnetic spin control. 

\small
\vspace{2cm}\noindent{\bf Methods}\\

\noindent{\bf ESR apparatus and sample configuration} The \HoW crystal was mounted in a 1.6~mm outer diameter ESR quartz tube and inserted between parallel metallized plates separated by 2~mm, which were used to apply \Efield pulses. Both the conductive plates and the sample were mounted in a standard Bruker MD5 resonator and could be rotated independently around the $x$-axis as shown in Fig.~\ref{fig1}b. By first studying the clock transition fields as a function of the sample rotation, we were able to determine the angle between the magnetic easy axis and the experimental rotation axis. In all experiments presented in this study, the magnetic easy axis was aligned as close as possible to $\boldsymbol{B}_0$. We applied the \Efield at a variable orientation $\varphi$ with respect to $\boldsymbol{B}_0$ (Fig. 1b).

Note that the magnetic easy axis is not coincident with any of the crystal facets or edges~\cite{Shiddiq2016,Ghosh2012}, nor is it collinear with the  pseudo-four-fold molecular symmetry axis, so it is not possible to guarantee the alignment by inspecting the crystal morphology.
Thus repeating the experiment on a different crystal gives a similar orientation dependence for the SEC, albeit the extrema of the SEC appear at different laboratory directions (see Ref~\cite{SI}, Section IV). This cosine-shaped orientation dependence of the SEC is indicative of the existence of an axial-type SEC in \HoW. The differences in behaviour between the two crystals arise from the variation in orientation of the crystals in the apparatus, and therefore the plane through which they are rotated in the laboratory frame (see Ref~\cite{SI} Section V for more details).

\noindent{\bf Microwave pulse sequences} A standard two-pulse Hahn-echo sequence ($\pi/2$ -- $\tau$ -- $\pi$ -- $\tau$ -- echo) was employed to measure an electron spin echo on the \HoW crystals. Short broadband pulses, 32~ns for $\pi/2$-pulses and 64~ns for $\pi$-pulses, were used for measuring the SEC parameters (data shown in Figs.~\ref{fig1} and \ref{fig2}). These pulse parameters were adjusted in order to optimise the echo intensity. By contrast, long selective pulses, 400~ns for $\pi/2$-pulses and 800~ns for $\pi$-pulses, were applied in the selective spin manipulation experiments (Fig.~\ref{fig3}). These pulse lengths were chosen to excite a narrow frequency population of spins, which can be selectively manipulated using the \Efield-induced frequency shift, while still giving a reasonable signal-to-noise ratio in spin echo measurements.

\noindent{\bf Calculation of SEC} The \Efield-induced change in the clock transition frequency is evaluated by computing the evolution of the molecular electronic structure using the multireference Complete Active Self-Consistent Field Spin-Orbit (CASSCF-SO) method (implemented by OpenMOLCAS~\cite{fdez2019openmolcas}) with the combined effect of the crystal field and the spin-orbit coupling calculated using SINGLE-ANISO module~\cite{ungur2017ab}. The effect of the \Efield on the easy axis orientation was also estimated (for details, see Ref~\cite{SI} Section VIII). The theory study was performed using both the crystal structure obtained and the relaxed structure optimised at the DFT level.


\noindent{\bf Acknowledgements}\\ 
This work is supported by: the EU (ERC-2014-CoG-647301 DECRESIM, ERC-2018-AdG-788222 MOL-2D, COST Action CA15128 MOLSPIN, the QuantERA project SUMO, and the H2020 research and innovation programme projects SPRING (No 863098) and FATMOLS (No 862893)); the Spanish MINECO (grant CTQ2017-89993 co-financed by FEDER, grant MAT2017-89528; the Unit of excellence ``Mar\'ia de Maeztu'' CEX2019-000919-M); the Generalitat Valenciana (Prometeo Program of Excellence); and the UK EPSRC (EP/P000479/1). J.J.B.\ acknowledges support by the Generalitat Valenciana (CDEIGENT/2019/022). J.M.\ is supported by Magdalen College, Oxford. J.L.\ is supported by the Royal Society through a University Research Fellowship.

\noindent{\bf Author Contributions}\\ 
J.L., E.C., A.G.-A.\ and A.A.\ conceived the study. Materials were synthesised by Y.D.\ under the supervision of E.C. ESR experiments were conducted by J.L.\ and J.M. Data analysis was performed by J.L. with input from A.A. Theoretical modelling was done by A.U., assisted by J.J.B, guided by A.G.-A., and in discussion with J.L., E.C.\ and A.A. All authors contributed to the manuscript.

\noindent{\bf Competing Interests}\\
The authors declare no competing interests. 

\noindent{\bf Additional Information}\\ 
Supplementary Information is available for this paper. 


\begin{thebibliography}{10}
\bibitem{Enote}
\bibinfo{note}{An oscillation in the quadrature component can also arise if the
  $E$-field pulse is accompanied by a parasitic magnetic field pulse that
  distorts $\boldsymbol{B}_0$; it contributes a trivial phase to the echo via
  the Zeeman interaction}.
\end{thebibliography}

\end{document}